\documentclass[10pt,a4paper]{article}
\usepackage[utf8]{inputenc}
\usepackage{amsmath}
\usepackage{amsfonts}
\usepackage{amssymb}
\usepackage{graphicx}
\usepackage{enumitem}
\usepackage{mdwlist}
\usepackage{pdfpages}
\author{James Lockwood, Susan Bergin}
\title{A neurofeedback system to promote learner engagement}
\date{March 2016}

\begin{document}
	
	\begin{titlepage}
	 \begin{figure}[ht!]
	 	\centering
	 	\includegraphics[width=.9\linewidth]{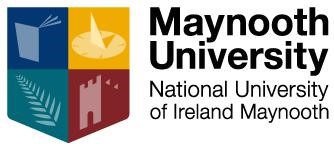}
	 \end{figure}
		 
	\begin{center}
		\Large\textbf{A neurofeedback system to promote learner engagement\\}
	\end{center}
	
	\begin{center}
		\Large\textit{James Lockwood, Susan Bergin\\}
	\end{center}
	 \hspace{0.5cm}
	 \begin{center}
	 \Large{Department of Computer Science,\\ 
	 	Maynooth University,\\ 
	 	Maynooth, Co. Kildare,\\ 
	 	Ireland}
	 \end{center}
	
	  \begin{center}
	  	Date: March 2016
	  \end{center}
	 \hspace{1cm}
		\paragraph{Abstract}
	    
		This report describes a series of experiments that track novice programmer’s engagement during two attention based tasks. The tasks required participants to watch a tutorial video on introductory programming and to attend to a simple maze game whilst wearing an electroencephalogram (EEG) device called the Emotiv EPOC. The EPOC’s proprietary software includes a system which tracks emotional state (specifically: engagement, excitement, meditation, frustration, valence and long-term excitement). Using this data, a software application written in the Processing language was developed to track user’s engagement levels and implement a neurofeedback based intervention when engagement fell below an acceptable level. The aim of the intervention was to prompt learners who disengaged with the task to re-engage. The intervention used during the video tutorial was to pause the video if a participant disengaged significantly. However other interventions such as slowing the video down, playing a noise or darkening/brightening the screen could also be used. For the maze game, the caterpillar moving through the maze slowed in line with disengagement and moved more quickly once the learner re-engaged. The approach worked very well and successfully re-engaged participants, although a number of improvements could be made. A number of interesting findings on the comparative engagement levels of different groups e.g. by gender and by age etc. were identified and provide useful pointers for future research studies.
	
	\end{titlepage}

	\section {Introduction}
	This project focussed on tracking student engagement levels during cognitive attention-based tasks. This report will provide scientific research on electroencephalogram (EEG0 machines and how they can be used to track engagement levels as well as discuss the use of self-paced tutorial videos as learning support tools. It will also describe the development of a series of experiments which use the EPOC to track student’s engagement whilst they take part in two purpose built cognitive tasks. These experiments include a neurofeedback intervention to encourage participants who dis-engage from the given tasks to re-engage. This project aims to show that neurofeedback can be used to re-engage learners during cognitive tasks.
\\\\
	Student learning and assessment are primary concern for educational institutions. Methods to improve student engagement and learning are a constant topic of discussion at learning and engagement conferences for example at ICEP and LIN, annual Irish Pedagogical conferences. Recently learning videos have gained acceptance as a valuable enhancement to improve student learning (including in Maynooth University’s Computer Science Department). However, do these videos engage students? Can an EEG device such as the EPOC allow a system to be developed to prompt a student if they do dis-engage and cause them to re-engage? This project will attempt to answer these questions.

	\section{Background}

	A review of the literature was carried out on (1) the relationship between engagement and learning, (2) the use of electroencephalogram (EEG) devices (especially the Emotiv EPOC used here) for monitoring neural activity during cognitive tasks, and (3) the use of videos as a learning tool.\\\\
	An EEG is a non-invasive technique to measure activity in the brain. The EPOC is a 14 channel EEG device for research. It is a dry (uses a small amount of saline) high resolution consumer grade EEG system. A number of verification studies have been carried out comparing the EPOC to a clinical EEG but There has been little independent study as it is a relatively new technology. However, proprietary claims suggest its appropriate for research and, of the limited studies, all have found that it performs well and can be used in place of more expensive units. Taking this into consideration the results obtained can be considered accurate for example, Ekanayake (2010) concluded that the EPOC does capture real EEG data but has considerable noise. This noise can be minimised using different techniques such as averaging It was also found that electrode placement is fairly fixed and thus this reduces the number of studies it can be used for. Andujar (2011) used the EPOC in a study that compared the engagement level of participants studying a paper handout with using a video game to learn about the Lewis and Clark Expedition (www.classbraingames.com). The game disseminated information about the expedition and the control group received the same information in the handout using only text.  Twenty-six participants took part (13 in the experimental group and 13 in the control). Each group had 20 minutes to complete the learning task. Participants wore the EPOC during the task and engagement levels were measured. The results of the study suggested that educational video games might not be significantly engaging and found that learning from handouts could be better for retaining information. This is an interesting finding and the closest study we could find on the use of the EPOC to measure engagement. But this study uses a video game and not a video tutorial and it could be argued, that a tutorial video reduces any cognitive load associated with aspects of the games or story-line issues which might impact on learning.\\\\
	Aside from EEG studies on engagement other empirical studies on engagement have highlighted its importance for learning and comprehension, some examples follow. Carini et al. (2006) found that student engagement is linked positively to desirable learning outcomes such as critical thinking and grades. In a study on the use of clickers in the classroom Blasco-Arcas (2013) found that increased engagement through the clickers improved student learning. Other studies have shown that increased engagement in reading leads to higher comprehension (Guthrie \& Wigfield, 2000 and Miller \& Meece, 1999).\\\\
	With respect to using educational video to improve student learning, Merkt et al. (2011) conducted two complementary studies, one in the laboratory and one in the field, comparing the usage patterns and effectiveness of interactive videos and illustrated textbooks when German secondary school students learned complex content. They used two separate videos of differing degrees of interactivity (the second of which they made themselves) and an illustrated textbook. Both studies showed that the effectiveness of interactive videos was at least comparable to that of print, in contrast to previous studies working with non-interactive videos,.  More recently, Willmot et al. (2012) show that there is strong evidence that creating an on-going video report during projects can inspire and engage students when incorporated into student-centred learning activities through: increased student motivation, enhanced learning experience, higher marks, potential for deeper learning of the subject, enhanced team working and communication skills, and by providing a source of evidence relating to skills for interviews. These studies indicate that videos can be used as an enhancement to a lesson, or unit of study. Furthermore, Guo et al. (2014) provided a list of recommendations on what makes a video engaging and beneficial. To do this they used both data analysis of 6.9 million video watching sessions and interviews with video production staff.
	These recommendations and findings include: shorter videos are much more engaging, videos where instructors speak relatively fast and with high enthusiasm are more engaging and that videos should incorporate motion and continuous visual flow in tutorials, along with extemporaneous speaking. Pre- and post-production planning is essential as topics can be segmented into smaller sections. 
	They also argue that filming in an informal setting is more engaging than a recording studio. The recommendations provided above influenced the development of the threshold video described in Hegarty-Kelly et al. (2015).
	\\\\
	The goal of the work presented here is to monitor, track and respond to changes in engagement whilst cognitive tasks are being attended to by gathering neuro-physiological data from the student. A neurofeedback system like this, which can detect, track and respond to these changes in an online learning environment could be beneficial to learning, especially online learners in the absence of a teacher to monitor and gauge attention. This is especially true and timely with the large growth of MOOC’s (Massive Open Online Courses) in recent years.\\\\
	This paper describes a neurofeedback system which monitors and responds to engagement levels in two cognitive tasks. First, whilst watching a tutorial video the feedback system would pause the video if a learner disengaged and only restart after re-engagement had occurred. This system could be of huge benefit in MOOC’s and long-distance learning courses which use videos as a major learning tool. The second task is a maze game in which a caterpillar moves around a maze, and the speed at which this happens is related to a participant’s engagement level. This application has the potential to be used in therapy settings for conditions such as Attention deficit hyperactivity disorder (ADHD). This engagement therapy taps into the brains ability to re-wire its own connections (neuroplasticity) and has been shown to be a superior method of treatment for ADHD than classical medicated interventions of the past (Fuchs et al. 2003).

	 \section{Method}
	 \subsection{Hardware and software description}
	 The Emotiv EPOC was designed for practical research applications and comes with a software suite that measures several emotional states including engagement, as well as boredom, excitement, frustration and meditation level in real time. Two applications were developed. The first monitored a user watching a video whilst wearing the EPOC and allowed interventions to take place depending on a user’s engagement level. The second was a maze game, where an object would travel around a maze and the speed at which it moved was determined by the user’s engagement level. The idea for this game was based on similar neurodfeedback games developed for children with ADHD such as that of Fuchs et al. (2003). Their version of pac-man was developed for children with ADHD where if they start to lose focus on the game pac-man starts to fade. The idea of the maze game developed here, in the Processing language (Foundation, 2016), was a very simplistic idea in which a caterpillar-like object moves around a maze. The speed of the caterpillar is related to the participants level of engagement, the higher the engagement level the faster the caterpillar moved, the lower the the engagement, the slower it moved.
	 \\\\ 
	 \begin{figure}[ht!]
	 \centering
	 \begin{minipage}{.5\textwidth}
	 	\centering
	 	\includegraphics[width=.9\linewidth]{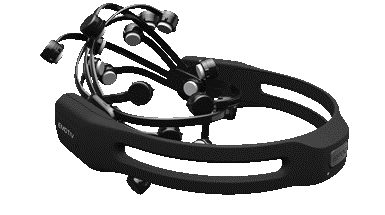}
	 	\caption{EPOC EEG headset\label{overflow}}
	\end{minipage}%	
		 	\begin{minipage}{.5\textwidth}
	 		\centering
		 		\includegraphics[width=.9\linewidth]{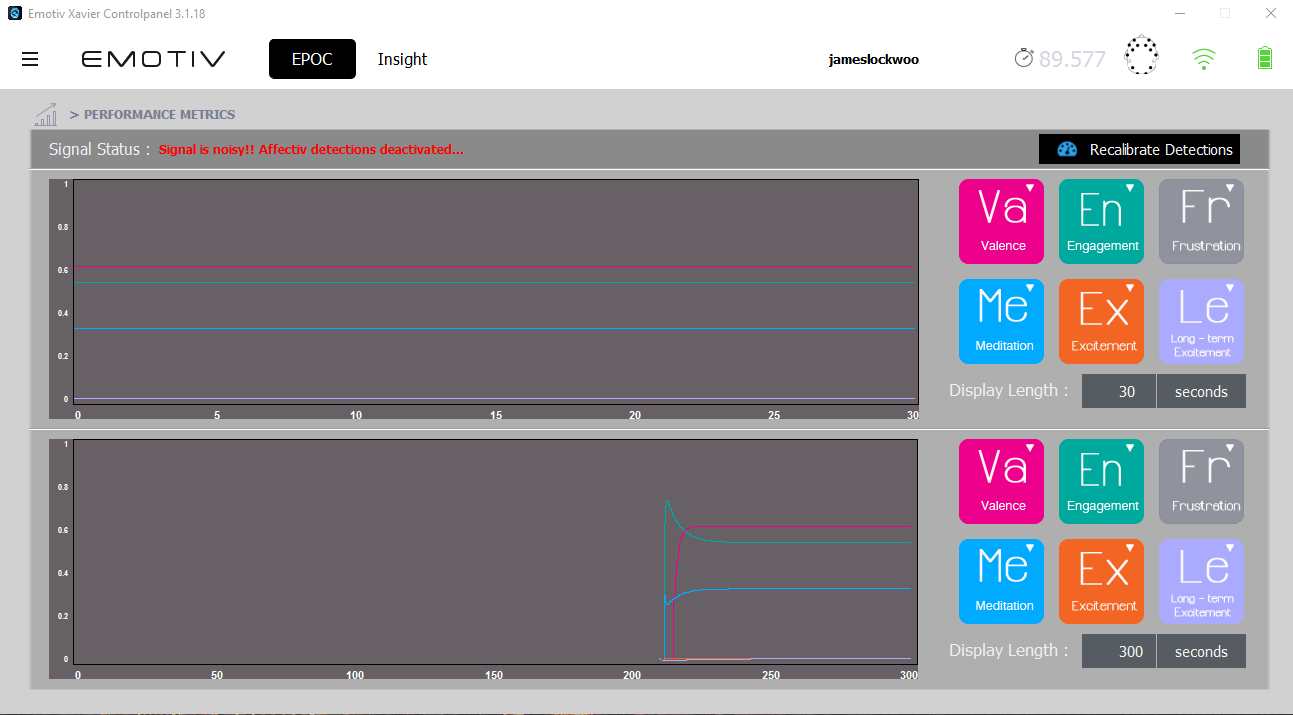}
		 		\caption{Emotiv Control Panel – Performance Metrics and Emotional States Suite \label{overflow}}
		 	\end{minipage}	
		 \end{figure}

	 \subsection{Participants and pilot study}
As this study included human particpants ethical approval was sought and granted for the study. All participants signed a consent form and an information sheet was provided prior to the session with questions. In total 21 final year students along with 5 Postgraduate students were recruited to participate in the study (16 of the final year and 4 of the Postgraduates students studied CS). A reasonable gender representaiton was sought.  A video of 8 minutes duration on if statements, was  used. This is material that all 4th year CS students would be extremely familiar with and so would potentially increase the likelihood that the participants would disengage and show the capabilities of the system. 
	
	 \subsection{Instruments}
	 A post-experiment questionnaire was used to gather feedback on perceived engagement for both the maze and video tasks based on a survey by Wigfeld \& Guthrie (1997) (further adapted by Hyun-Gyung Lee (2012)), which all participants filled out directly after doing both the video and maze tasks (see Appendix A). All data was anonymised and each participant was given a unique identifying code. This code was used by participants in another survey that gathered background information including gender, age, qualification level, programming experience etc. 
	 \subsection{Experimental Design}
	
	 This survey was filled out prior to the experiments taking place. All participants took part in all of the tasks in the following order: Calibration, Video, Video Survey, Maze and Maze Survey. Although it can be preferential to counter-balance an experiment that involves two discrete components, it was decided to always show the video first as this is of primary importance to us.
	 Before the main study a three-person pilot study was carried out to ensure that each of the experiments worked correctly and to test whether the selected baseline engagement value (set at 0.5, the mid-point value for engagement in the proprietary software) was reasonable. After two participants it was clear that this crude approach to measuring engagement was insufficient and an individualised calibration procedure was needed as engagement levels differ greatly from person to person. As such a calibration process was carried out where (1) the participant fixated on a black cross on a grey screen for sixty seconds, and (2) the participant closed their eyes for sixty second. During the latter, participants were instructed to try to relax and to not focus on anything in particular. Whilst their eyes were closed the program took a running average of their engagement level for that minute and this value was used as each particpants baseline value. In theory, when the participants eyes are closed they should have been disengaged and this was confirmed in their levels. This baseline was then used as the threshold for whether the video should pause and also as to how slow the caterpillar would move in the maze game. This means that the video would pause and would not restart and, similary, the caterpillar would continue to go slowly, until the level was above the threshold value. The formula used for the caterpillar was as follows: if the engagement level was above the threshold the caterpillar moved at 10*Engagment level +1 , if it was below this level then it moved at 1*Engagment level + 0.1. After the application had been designed, a third and final participant took part in the pilot study; all three experiments were conducted on this participant. This calibration task was found to work well and interacted correctly with both the maze and video applications, however it is a simplistic technique which could use further testing and honing to correctly judge its effectiveness.
	 
	 \subsection{Experiments}

	 After building the experiments and running the pilot study, the next step was to carry out a main study using the developed applications. Nineteen CS students and six non-CS students participated in the study. This resulted in twenty usable data sets (16 CS, 4 non-CS). On average the experiment took between 30-40 minutes to complete and the experiments were run over four weeks. Below is a full demographic breakdown of the participants from who data was successfully collected.	 \\\\
	 \begin{figure}[ht!]
	 	\centering
	 	\includegraphics[width=.9\linewidth]{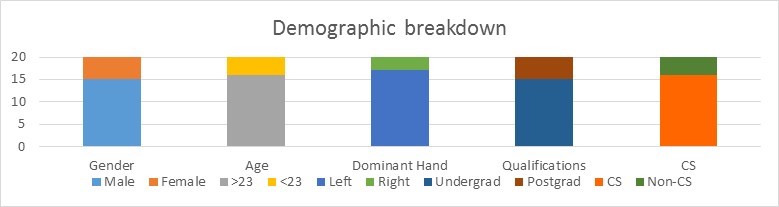}
	 	\caption{Breakdown of some of the demographics of the participants \label{overflow}}
	 	
	 \end{figure}

	 The experiments were run on a Lenovo IdeaPad U330p laptop (8GB RAM, Intel i5 processor running Windows 10) which participants had on a table in front of them. The first author remained in the room to monitor the headset’s signal as occasionally it would drop out, that is, one or more of the contacts would move out of ideal placement. After reading the information sheet the participants were instructed, and helped as appropriate, on the correct placement of the EPOC on their heads. Once a good signal was achieved (i.e. the 14 channels were detecting a good signal) participants took part in the calibration procedure. After an individual baseline had successfully been obtained through the calibration system, they then viewed the video, being advised to watch the video and that if it paused, to try to refocus on the task. Noise and distractions were limited as phone calls, emails etc. were not able to be received in the room, the room was also quiet with little internal/external noise. After completing the video, the participants were asked to fill out an engagement questionnaire. If needed, extra saline solution was applied to the sensors of the EPOC. The participants then completed the maze task and again filled out an engagement questionnaire.
	 Twenty full data sets were successfully obtained over a period of four weeks. Six more participants took part but their data was either corrupt, too noisy or not obtainable. Four of these were female participants and the cause was most likely the amount and thickness of their hair prevented the EPOC’s sensors establishing a good connection. Another contributing factor may have been head size, that is, they had slightly smaller heads than other participants and so sensor contact may have been sub-optimal, this is down to the restricted movement of the sensors and flexibility available of the EPOC compared to medical EEG devices.

	 \section{Evaluation}
	 \paragraph{Data Analysis}
	 After collecting all the data statistical analysis was performed on the collected data. This included frequency tests, t-tests and P-values. Due to the small sample size, and even smaller breakdown, anything with a p-value of less than 0.15 (i.e. Alpha = 0.85) was treated as potentially important.\\\\
	 A point worth noting is that the uninteresting nature of both the video content and the production quality were almost unanimously confirmed in post-video response forms (19/20 participants responded to the question “What did you find that was boring or too easy?” (participants were instructed to answer this question and all others in regards to the task completed, either the video or maze) with comments that the video itself or the content was boring or easy). At the beginning of each experiment participants were told what would occur if they disengaged, but were not informed to try harder than normal to engage, just to watch the video/maze and try to re-engage if the video paused or the caterpillar slowed.\\\\
	 Findings from statistical analysis of the data are provided next. Although the sample size is small (n=20), some of the findings are significant enough to warrant further investigation.
	 
	 \subsection{Calibration}
	 	 
	 	\paragraph{Finding 1 – During the calibration task, participants with a lower baseline engagement level during eyes shut had higher engagement in the video and the maze game tasks.} The most significant finding (throughout all three tasks) was the fact that those who had lower engagement levels with their eyes shut (the value taken as a baseline) caused the video to pause less and had similar results with the maze game i.e. caused it to move slowly for less of the time. To compare the two groups, the average of the whole populations baselines (worked out at 0.501407) was taken and the group was then split using this value. The baseline values were the unique values that were calculated for each participant during the calibration process.\\\\
	 	All participants with a baseline below this population average (n = 9, referred to here as the Low Group (LG)) were then compared with those above it (n=11 referred to as High Group (HG)). When comparing the LG’s average engagement levels when fixating on the cross, they had statistically lower scores than the HG (p-value = 0.017) during the same period. This group also had a larger range on these values (p-value = 0.018). This is interesting as it suggests those who are able to disengage before completing the tasks are more likely to succeed at the cognitive task. This could help to show that learning meditative techniques and applying these before a cognitive task (test, studying etc.) can improve performance (Zeidan et al., 2010).
	 	\paragraph{Finding 2 – During the calibration task, participants who wear prescription glasses had lower average engagement levels during the fixation cross section} Participants who wear prescription glasses (n=6) had a statistically lower average engagement level during the minute in which they focused on a fixation cross than those who don’t (p value = 0.148). These same participants also had a greater range in their average values (eyes open/shut) than those who don’t wear glasses (p-value = 0.131).  This tendency towards an effect is interesting but may not be found with a larger population. However, it may suggest something interesting such as, for example, that glass-wearers have learnt to concentrate harder when instructed than non-glass wearers. Such a hypothesis would have to be tested for merit however.
	 	\paragraph{Finding 3 – Participants over the age of 23 had a lower average baseline during the fixation cross section} It was also found that mature participants (over 23) had a significantly lower average baseline engagement level when staring at the fixation cross than those who are younger than 23 (p-value = 0.115). This could be an interesting finding given that mature participants seem to disengage with both the video and maze much less than non-mature participants (see below).
	 
	 	\paragraph{Finding 4 – There was significantly different ranges of engagement levels and the baselines (eyes shut section) between those who identified as morning/night people or those who gave no preference} Another interesting finding was that those participants who identified as morning people (n=5) and those who had no preference to morning or night (n=6) had significantly different ranges in their averages (eyes open/shut) than those who identified as night people (p-values = 0.019 in both cases). This is added to the fact that those who identified as morning people had on average a higher engagement level than those who identified as night people or gave no preference during the eyes shut section (p-values = 0.019 and 0.146 respectively).\\\\ 
	 	People with no preference seemed to have a lower engagement level than the two other groups when they had their eyes shut. This would have contributed to the range of their values however the statistical significance of this is not particularly strong (p-value = 0.217 against morning people and p-value = 0.254 against night people).\\\\
	 	It should be noted that participants who identified as night people (n=9) all took the experiments between the hours of 1pm and 7pm. This ensured that they didn’t take the experiments at an hour that could have diminished their cognitive ability. The same is true with participants who identified as morning people (n=5) who all took the experiment between 12pm and 3pm, so again any preference has been accounted for. 
	 	\paragraph{Finding 5 – Those with previous programming experience had significantly higher ranges than those who had no previous experience} A last and significant finding made from the calibration data was that those who have previous programming experience had a larger range (eyes closed compared to eyes open) on average than those with none (p-value = 0.057) during the calibration task. They also had a higher average level with their eyes shut, however this was not as statistically significant (p-value = 0.275). This could mean that they are not as good at disengaging having been looking at a computer screen and this in turn could be due to the amount of time spent at a PC is potentially increased through a CS degree program. However, this conclusion is contradicted slightly by the results of those who rated how often they programmed as 4-5 (1 being “Not at all”, 5 being “Daily”) had a lower average engagement level during the eyes open section compared to those rated as 2-3 (p-value = 0.228).
	
	 \subsection{Video Data}
	 With regards to the video related data initial analysis was focused on the number of times the video paused during the duration of the video. This indicated a drop below the baseline level recorded during the calibration task, referred to as a “drop” or “dropping” in the next few sections. Of the 20 participants, 7 watched the whole video without it pausing once. Of these seven participants, a brief summary of their breakdown is follows: 1 female; 6 males, 2 mature, 5 non-mature, 2 postgraduate students, 5 undergraduates, 1 non CS student, 6 CS students.
	 Across the whole population the mean number of drops was 12.55, the median was 9 and the mode was 0, the maximum value was 46 (see Figure 6). 
	 	\begin{figure}[ht!]
	 		\centering
	 		\includegraphics[width=.9\linewidth]{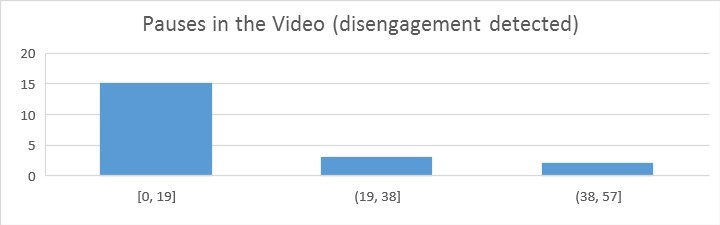}
	 		\caption{The number of pauses(drops) during the video playback \label{overflow}}
	 		
	 	\end{figure}
	 Some points of interest from this data include the following:
	
	 	\paragraph{Finding 1 – Participants with lower baseline engagement levels paused the video significantly less than those with higher baseline levels} As found with the calibration, participants whose engagement baseline was below the population average (0.501407) dropped significantly less than those who’s baseline was above this average. On average those who were below (n=9) paused 2.333 times whereas those above paused 20.9 times (p-value = 0.002). This is significant as it means those with a lower baseline level were more engaged with the video. This could support evidence that learning meditative can improve performance (Zeidan et al., 2010). Further investigation in this area is warranted.
	 	\paragraph{Finding 2 – Participants who rated their Java level as lower paused significantly more than those who rated their level as high}Another interesting statistic is that people who rated their Java level as 1-3 (n=10) (with 1 being “Very poor” and 5 being “Very good”) paused significantly more than those who rated it 4-5 (n=10). On average those who rated their Java level as lower paused 18.1 times with those rating it higher paused on average 7 times (p-value = 0.852). This is interesting as the video is designed for beginner programmers, and so in theory those with greater knowledge and experience of the language “should” find it less engaging. However most of the participants responded to the video survey negatively and found it boring so this could be down to the video itself being boring and poor production quality (see text analysis section below). This statistic also holds significance if you remove those who have never programmed before, for four of the participants this was true and they rated their level as 1 (no-one rated their level as 2). If you take those participants (n=6) who rated their Java level as 3 and compare it to those rated 4-5 then again those rated 4-5 paused significantly less, on average 19.33 times for level 3 participants compared to 7 for 4-5 (p-value = 0.125). It would be interesting to take this experiment and run it on those learning to program (possibly in two groups, one who want to continue CS and one just making up credits) to see if this holds for the target population. 
	 	\paragraph{Finding 3 – Mature participants (over 23) caused the video to pause less}One final point from the video data was that mature participants (n=4) caused the video to pause less on average than non-mature. Although not statistically different (p-value = 0.362, note very small sample size) it was numerically quite different, with matures averaging 6.5 drops compared to 14.06 with non-mature. Further investigation in this area is warranted.
	 	\paragraph{Finding 4 – No difference in people whose self-perceived engagement level was low (1-2) or high (3-5)}On another note, no statistical difference was found between those who responded to how engaged they were during this task with a 1-2 rating or a 3-5 rating (with 1 being “Not at all” and 5 being “Completely/Always”). Those who rated how engaged they were as 1-2 dropped on average 11.4 times compared to 13.7 times with those who rated engagement as 4-5 (p-value = 0.732). This is interesting as it could mean that people felt they were more engaged than they actually were, or conversely that they were less engaged than they felt. From anecdotal evidence and observation, many participants felt they were focusing and engaging in the video but still caused it to pause, however all agreed the video was boring and unengaging. This could mean that even though they felt they were engaging, the EPOC and associated baseline engagement level did not indicate they were.
	
	 	\subsection{Maze Data}
	 	During the maze game if participants’ engagement levels dropped below their baseline level, the speed of the caterpillar dropped dramatically. The speed was constantly changing due to the level but dropped drastically if this occurred. This is again referred to as a drop. 
	 	Of the 20 participants, 7 played out the duration of the game (3 minutes) without it dropping once. Of these seven, here is a brief summary of the breakdown: 7 males, 3 mature, 4 non-mature, 3 postgraduate students, 4 undergraduates, 7 CS students. It is also worth noting that 5 of the 7 who didn’t drop during the maze task also didn’t drop during the video task.
	 	Across the whole population the mean number of drops was 3.15, the median was 3 and the mode was 0, the maximum value was 10 (see Figure 7).
	 	\begin{figure}[ht!]
	 		\centering
	 		\includegraphics[width=.9\linewidth]{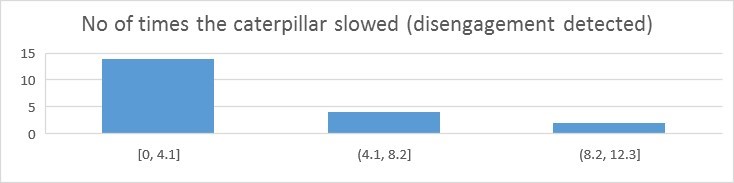}
	 		\caption{The number of times the caterpillar slowed (drops) during the maze game \label{overflow}}
	 		
	 	\end{figure}
	 	\paragraph{Finding 1 – Those with higher baseline engagement levels dropped (slowed the caterpillar) less than those with lower baselines}As previously noted, a significant difference was found in the number of drops in the participants whose baseline was below the population average. Those below the population average dropped (or slowed) on average 1.11 times compared to 4.81 times for those above the average (p-value = 0.021). 
	 		\paragraph{Finding 2 – Those who rated their Java level as low (1-3) dropped more than those who rated their level as high (4-5)} Also, similar to the video data, those who rated their Java level as low (1-3) caused the caterpillar to slow more times than those who rated it high (4-5). On average this was 4.3 times compared to 2 times (p-value = 0.126). It is worth noting that those who rated their Java level as 1 (n=4, all the non-CS participants) had the highest average drops in the maze with an average of 5, compared to 3.83 with level 3 (6/20), 2.5 with level 4 (8/20) and 0 with level 5 (2/20). This is interesting and warrants further investigation to identify why this is observed.
	 		\paragraph{Finding 3 – Mature participants (over 23) dropped less than those under the age of 23}Another similarity with the video data, with greater statistical significance, is that mature participants cause the caterpillar to slow less than non-mature. Matures causes it to slow on average 1 time compared to an average of 3.69 for non-mature (p-value = 0.149).
	 		\paragraph{Finding 4 – No statistical difference in those who found the maze more engaging, interesting or challenging than those who did not}In line with the data collected from the video task there was no statistical difference between participants who found the task more or less interesting, more or less challenging or more or less engaging.
	 	
	 \subsection{Maze vs Video}
	 In terms of survey feedback for both the maze and video, participants were asked to rate three survey items from 1-5 (see Appendix A for the descriptions associated with these values), they were as follows: (1) I was challenged by this task, (2) This task was interesting to me, (3) This task engaged me.
	 All participants took the maze challenge after the video and after having filled out the video feedback form.
	 What is interesting is that almost all participants found the maze more challenging, engaging and interesting than the video (see Figure 8).	 
	 	\begin{figure}[ht!]
	 		\centering
	 		\includegraphics[width=.9\linewidth]{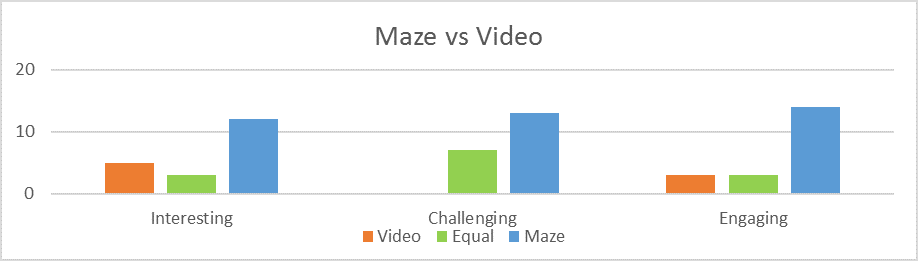}
	 		\caption{Number of people who rated the maze/video more or less in each question. \label{overflow}}
	 		
	 	\end{figure}
	 Below is an analysis of the responses (from 1-5, with 1 being Not at all and 5 being Completely/Always) that participants gave the three above questions which gives further weight to the claim that the maze game was more challenging, interesting and engaging to the participants.\\\\
	 For the Video responses the means of the three questions (referred to now as Challenging, Interesting and Engaging respectively) were: Challenging = 1.6, Interesting = 1.8 and Engaging = 2.5. The respective modes were 1, 1 and 3, the medians 1.5, 1 and 2.5 respectively and the ranges were Challenging = 2, Interesting = 3 and Engaging = 4.\\\\
	 For the Maze responses the mean of the three questions (referred to the same as above were: Challenging = 2.8, Interesting = 2.2 and Engaging = 3.35. The respective modes were 3, 1 and 4, the medians 3, 2 and 3.5 respectively and the ranges were Challenging = 4, Interesting = 3 and Engaging = 3.
	 \subsection{Text Analysis}
	 Text mining was carried out using R, however due to the small amount of text provided in the open questions on the response form, very little was found. For the video forms the words ‘video’, ‘interesting’, ‘focus’ and ‘boring’ were most frequent (37, 10, 10, 6 times respectively). ‘Interesting’, however usually relates to the headset or the fact their emotions were being tracked, only one participant stated that the video material was interesting and they also said that it became less interesting as it went on, this participant was a non-CS student. Another thing to note is that people generally felt they were focusing, or at least attempting to focus on the video, even if they were deemed disengaged by the EPOC and the video subsequently paused. This could be down to the crude metric for pausing the video. A more sophisticated algorithm to monitor and track engagement levels prior to pausing the video could lead to perceived engagement being matched more accurately with the system.\\\\
	 For the maze data the most common words were ‘speed’, ‘snake’, ‘focus’, ‘caterpillar’, ‘task’ and ‘maze’ (17, 13, 13, 11, 9, 9 times respectively). Sentiment analysis revealed that participants in general had a more positive response to the maze task than the video as indicated by their responses to the questions on the feedback form. Participants used different approaches to remain focused in the maze game, including following the caterpillar/snake, guessing where it would go and focusing just ahead of it, this is in contrast to the video where people struggled to focus. In the video responses people recalled attending to mistakes in the video and others thought of different things (not related to the video) so as not to appear disengaged. This latter finding is unsettling as the goal is to monitor engagement with the learning video not concentration but on a different matter. An eye movements study may help to automatically determine if this is the case.
	 \section{Conclusions and future work}
	 
	 This project has shown that development of an intervention system for reengaging learners can be done and is effective. An EEG machine can be used to track the learner’s engagement level and then this reading can be used to adjust the content being displayed, in this case pausing. This background work could be expanded into a more robust system which can improve learner’s experience and improve engagement overall performance.\\\\
	 The background work done on the EPOC, namely the testing of its capabilities will be a great benefit to both the Maynooth University Computer Science Education Research Group in further projects but could also benefit other groups that use or are considering using the EPOC in similar experiments.\\\\
	 The findings from the experiments also provide interesting insight into the use of videos, specifically the tutorial video, and how engaging they are for students. From the analysis the most significant findings that warrant further investigation are:
	 \begin{itemize}
	 	\item Those with lower baseline engagement levels caused both the video and the maze game to drop less than those with higher baseline levels. This could support evidence that practising meditative methods increases cognitive performance.
	 	\item Those who rated their Java level as low (1-3) caused both the video and the maze to drop more than those who rated their level as high (4-5). This is interesting especially given the nature of the video (a beginner Java tutorial) and it would be interesting to further investigate this with similar videos as well as unrelated media. 
	 	\item Mature participants (those over 23) caused the video and maze to drop less than those under the age of 23. This is another interesting finding as potentially it could show, with further and more intensive study, that those over the age of 23 may find it easier to engage with cognitive tasks than those younger. However, this is a purely hypothetical thought but even in isolation it is an interesting finding from this project.
	 \end{itemize}
	 As stated, given the small sample size, the findings, although interesting, warrant further investigation.\\\\ 
	 In the future it would be interesting to do a repeated measures test with a sample group where engagement levels were measured for both a “boring” tutorial video, a “good” tutorial video and an engaging video such as a commercial, film trailer, television clip etc. This could give a better indication on both what makes an engaging tutorial video (or doesn’t) and also how they fair compared to an entertainment video.  It could also be useful for building a sophisticated algorithm for determining individualised engagement levels.

	

\section*{Supplementary material}

# Appendix A

**Maze**                              **Participant code:**_____

## Section 1

| *Instructions:* **Circle one response for each item.** | Not at all | Partially/ Slightly | Somewhat | A great deal | Completely/ Always |
|---|---|---|---|---|---|
| This task was interesting to me. | 1 | 2 | 3 | 4 | 5 |
| I was challenged by this task. | 1 | 2 | 3 | 4 | 5 |
| I was interested in this task because others were interested in it. | 1 | 2 | 3 | 4 | 5 |
| This task engaged me. | 1 | 2 | 3 | 4 | 5 |
| During this task I thought about things not related to this task. | 1 | 2 | 3 | 4 | 5 |
| During this task I was aware of distractions. | 1 | 2 | 3 | 4 | 5 |
| During this task I controlled my learning | 1 | 2 | 3 | 4 | 5 |
| I could express myself freely during this task | 1 | 2 | 3 | 4 | 5 |

## Section 2

Please answer the questions below as completely and concisely as possible.

1. What in this task did you find challenging?

2. What did you find that was boring or too easy?

3. What was too hard or that you didn't have the appropriate skills or knowledge for?

4. What made this task interesting or not interesting to you?

5. What helped you focus on this task?

6. What, if anything, made you lose focus during this task?

7. What parts of the task did you have control over?

8. What parts of the task could you not control?

# Consent Form

| Research Project: | An investigation on the use of neurofeedback to improve learner engagement. |
|---|---|
| Aim: | Identify, monitor and respond to engagement during online learning activities. |
| Researcher: | Mr James Lockwood, Department of Computer Science, Maynooth University, Maynooth, Co. Kildare<br>Dr Susan Bergin, Department of Computer Science, Maynooth University, Maynooth, Co. Kildare |
| Contact details: | James Lockwood <james.lockwood.2013@mumail.ie><br>Susan Bergin susan.bergin@nuim.ie |

I, _____ agree to participate in the research project being carried out by Mr James Lockwood and Dr Susan to investigate the use of neurofeedback during online learning tasks. The research will be carried out at the Department of Computer Science, Maynooth University.

My participation will consist essentially in completing three tasks. The first task will require me to answer questions related to my engagement and to provide standard background information about myself. The second task will require me to watch a tutorial video and the third task will require me to play a simple game to move a caterpillar around a maze. Standardised non-invasive physiological measurements of my engagement whilst I am completing the tasks will be recorded.

The data gathered will only be used by the researchers above and the findings may be published in suitable conferences and journals. I can access my data at my discretion. I understand that the research will not form any kind of medical diagnosis or treatment.

I have received assurance from the researchers that the information that I will share will remain strictly confidential and that no information that discloses my identity will be released or published. However, I recognize that, in some circumstances, confidentiality of research data and records may be overridden by courts in the event of litigation or in the course of investigation by lawful authority. In such circumstances the University will take all reasonable steps within law to ensure that confidentiality is maintained to the greatest possible extent

I am free to withdraw from the study at any time until the work is published. I can refuse to answer any of the questions asked or to participate in any of the tasks. All data gathered will be stored in a secure manner and only the above named researchers will have access to it. There are two copies of the consent form, one of which I may keep. If I have any questions about the research project, I may contact the named researchers above at the contact details provided.

*If during your participation in this study you feel the information and guidelines that you were given have been neglected or disregarded in any way, or if you are unhappy about the process, please contact the Secretary of the Maynooth University Ethics Committee at research.ethics@nuim.ie or +353 (0)1 708 6019. Please be assured that your concerns will be dealt with in a sensitive manner.*

Name: _____

Signed: _____      Date:_____

Researchers Signature: _____      Date:_____

# Information Sheet

| Research Project: | An investigation on the use of neurofeedback to improve learner engagement. |
|---|---|
| Aim: | Identify, monitor and respond to engagement during online learning activities. |
| Researcher: | Mr James Lockwood, Department of Computer Science, Maynooth University, Maynooth, Co. Kildare<br>Dr Susan Bergin, Department of Computer Science, Maynooth University, Maynooth, Co. Kildare |
| Contact details: | James Lockwood <james.lockwood.2013@mumail.ie><br>Susan Bergin susan.bergin@nuim.ie |

### The Purpose:

The purpose of the research study is to investigate the use of neurofeedback during online learning tasks.

### The Participant:

The participant mentioned throughout this information sheet, refers to an adult who is currently participating a student at the Department of Computer Science, Maynooth University.

### The Tasks:

Participants will be required to complete three distinct tasks. The first task will require the participant to answer standard questions related to their background (age, gender, dominant hand, native speaker etc.) and their level of engagement. During the second task the participant will watch a tutorial video and during the third task the participant will play a simple game to move a caterpillar around a maze. In both these tasks standardised non-invasive physiological measurements of engagement will be recorded and monitored. To collect the physiological data a standardised made-for-purpose EEG headset will be worn by the participant.

### The Procedure

The experiment will take place in room 3.13, the RF shielded room on Level 3, Eolas Building, Maynooth University.

The participant will then be given an informed consent form and information sheet. The participant will be then reassured that all gathered data will be immediately encoded so that participants will only be identified by a reference number and no identifying information will be stored. The participant will be asked if they have any visual or reading difficulties that might interfere with the tasks. A participant must have normal or corrected-to-normal vision and ideally should be native English speakers or have a sufficient mastery of English to study in Ireland (as measured by a recognised English examination e.g. score of 6.0 UG or 6.5 PG IELTS examination).

Once consent has been given, the participant will be asked to complete a short background survey. The participant will then be asked to sit facing the computer. The participant will then be asked to wear the Emotiv EPOC EEG headset. The headset will be set up and calibrated, this process will take no longer than 15 minutes. The headset is a 14 channel (sensor) EEG head band. A small amount of saline solution will be added to each sensor to amplify the electrical signal from the brain that it will record. It is light and comfortable. Basic EEG signals will be measured to detect engagement. The headset is non-invasive and will not affect the participant is any way.

After the participant has read the instructions the researcher will monitor the headset status from a separate monitor. The participant should commence the task by pressing the spacebar.

### Non participation and exit from the study:

There is no requirement to participate in this study. A participant may choose to exit the study at any time up until the work is published. The participant must send their request in writing to the researcher and all data collected to date for that participant will be destroyed.

### Anonymity and security of data:

As soon as the data is collected it will be encoded with a unique identity key. No records of the participant's identity will be stored. It must be recognized that, in some circumstances, confidentiality of research data and records may be overridden by courts in the event of litigation or in the course of investigation by lawful authority. In such circumstances the University will take all reasonable steps within law to ensure that confidentiality is maintained to the greatest possible extent.

The data will be stored on a secure server located in Ireland. Only the above named researcher will have access to it. The data will be kept for the duration of the research and will be destroyed thereafter.

### Access to your data:

At any time, a participant can request a copy of their data that is stored by contacting the researcher. The participant will need to provide their unique key so that the researcher can identify their data.

### Questions:

If you have any further questions, please contact the researchers using the above contact details.

In the unlikely event that you experience any discomfort after the procedure please contact a doctor. The Maynooth University doctor is:

Dr Gaffney, Glenroyal Centre, Maynooth. Tel: 01 6291169

**Wave**
*Scientific*

Wave Scientific Ltd • 3 – 5 Vinalls Business Centre • Nep Town Road • Henfield • West Sussex • BN5 9DZ

# Report number: R0251

# Shielding Effectiveness Measurements
# of a shielded room at
# National University of Ireland, Maynooth

| | |
|---|---|
| Client name: | Global EMC (UK) Ltd. |
| Client address: | Prospect Close<br>Lowmoor Road Industrial Estate<br>Kirkby-in-Ashfield<br>Nottinghamshire<br>NG17 7LF<br>United Kingdom |
| Measurement site: | ICT Hub, NUI Maynooth<br>North Campus<br>Maynooth<br>Co. Kildare<br>Ireland |
| Purchase order number: | PO4906 |
| Our project number: | 30237 |
| Measurement date: | 13 November 2014 |
| Report Date: | 17 November 2014 |
| Measured by: | Nick Smith |
| Approved signatory: | |

N J Smith

# Contents

# 1. Measurement site details

The measurement site was a shielded room of approximate dimensions 5.8m x 2.8m x 3m high. The room was of a modular (pan type) construction and was fitted with filters, air vents, a penetration panel and a shielded door, as shown in figure 3.

# 2. Test specifications

Measurements were performed in general accordance with the following standard:

EN 50147-1 [1]

## 2.1. Measurement frequencies

Measurements were made at the following frequencies and field types:

| | |
|---|---|
| **1 GHz** | **Plane wave** |
| **5 GHz** | **Plane wave** |
| **10 GHz** | **Plane wave** |

# 3. Measurement methodology

Testing of shielding effectiveness (or attenuation) involves determination of the ratio of $V_0/V_s$, where $V_0$ is the voltage induced into the receiving antenna at a specified test distance from the transmit antenna. $V_s$ is the voltage induced when the shield is between the antennas at the same reference distance.

This ratio can be conveniently expressed in dB and the procedure for determining the shielding effectiveness is detailed below.

At each frequency, the shielding effectiveness, **S** is obtained from the result of two transmission loss measurements:

**$V_0$** is a reference level measurement and is the measured level (in dBuV) without any shielding between the transmit and receive antennas
**$V_s$** is the measured level (in dBuV) with the shielding between the antennas

The shielding effectiveness (expressed in dB) is then given by

**$S = V_0 - V_s + A_{att}$**

where **$A_{att}$** is the attenuation (in dB) of a calibrated coaxial attenuator that is inserted for the reference level measurement ($V_0$) in order to protect the input of the spectrum analyser from overload.

Details of a typical equipment configuration for the reference level measurement are shown in Figure 1. The equipment configuration for the shielding effectiveness measurement is shown in Figure 2. Amplifiers and pre-amplifiers are chosen depending on the test frequency and required measurement dynamic range.

The transmit antenna was positioned outside the enclosure and the receive antenna positioned inside the enclosure. The distance between the antennas was maintained the same as for the reference reading within practical restraints. A nominal measurement distance of 2.0 m distance was used for the plane wave measurements.

At each test position, the receive antenna was scanned around nearby joints and seams in order to determine the minimum shielding performance level. Where access to locate the transmit antenna was limited by the host building then these areas were assessed when measuring at nearby test points.

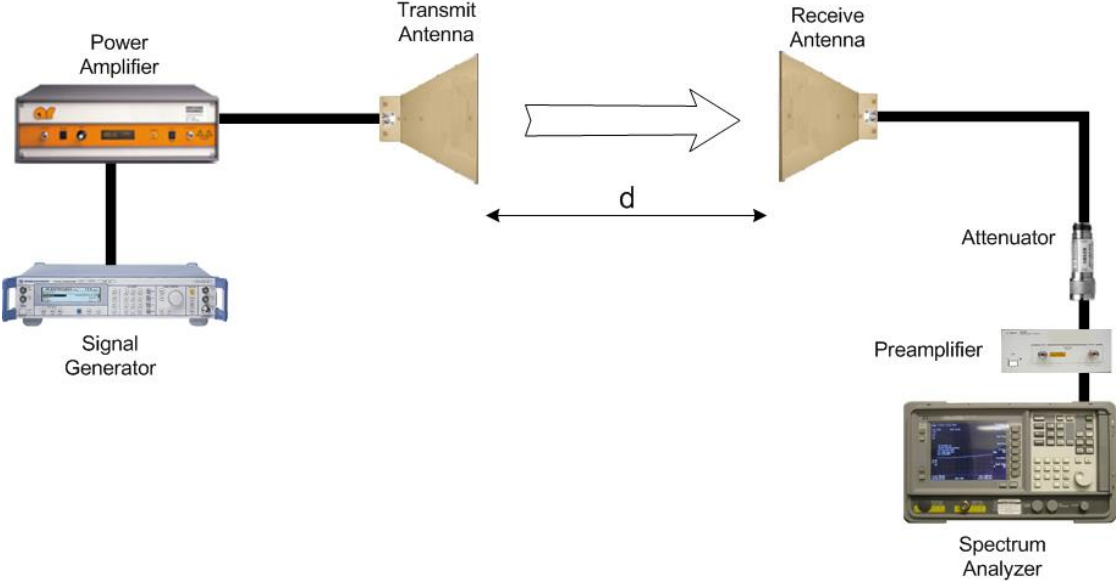

**Figure 1:** Typical test equipment configuration for the reference level measurement

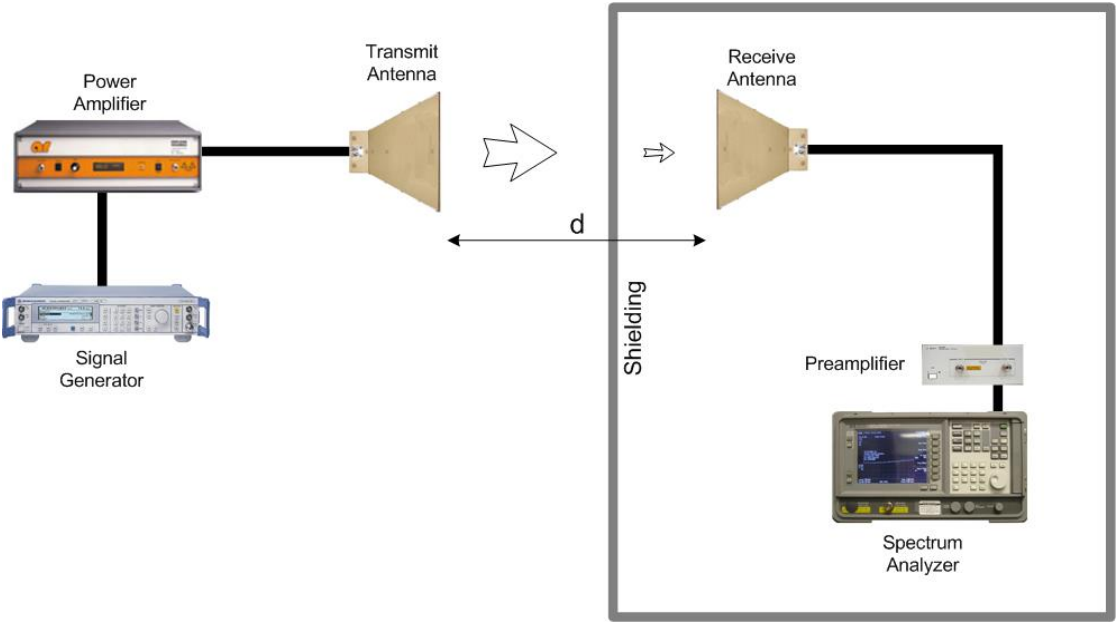

**Figure 2:** Typical test equipment configuration for shielding effectiveness measurement

# 4. Measuring equipment

The following equipment was used for the measurement of shielding effectiveness:

| Equipment | Serial Number |
|---|---|
| Hewlett Packard 8593E Spectrum analyzer | 3649A02614 |
| Rohde & Schwarz SMY02 signal generator | 832687/036 |
| Giga-tronics 7100/2-20 signal generator | 742818 |
| Schwarzbeck UHAL 9107 antenna | 1269 |
| Schwarzbeck UHAL 9107 antenna | 1409 |
| AH Systems SAS-571 horn antenna | 1252 |
| AH Systems SAS-571 horn antenna | 1253 |
| Wave Scientific 3W2G5 RF amplifier | 325001 |
| Logimetrics A340/CJ  TWT amplifier | 4280 |
| Wave Scientific WPA-15109A preamplifier, 100 kHz – 10 GHz | 1308002 |
| Precision fixed attenuators | N/A |

# 5. Shielded room and test point details

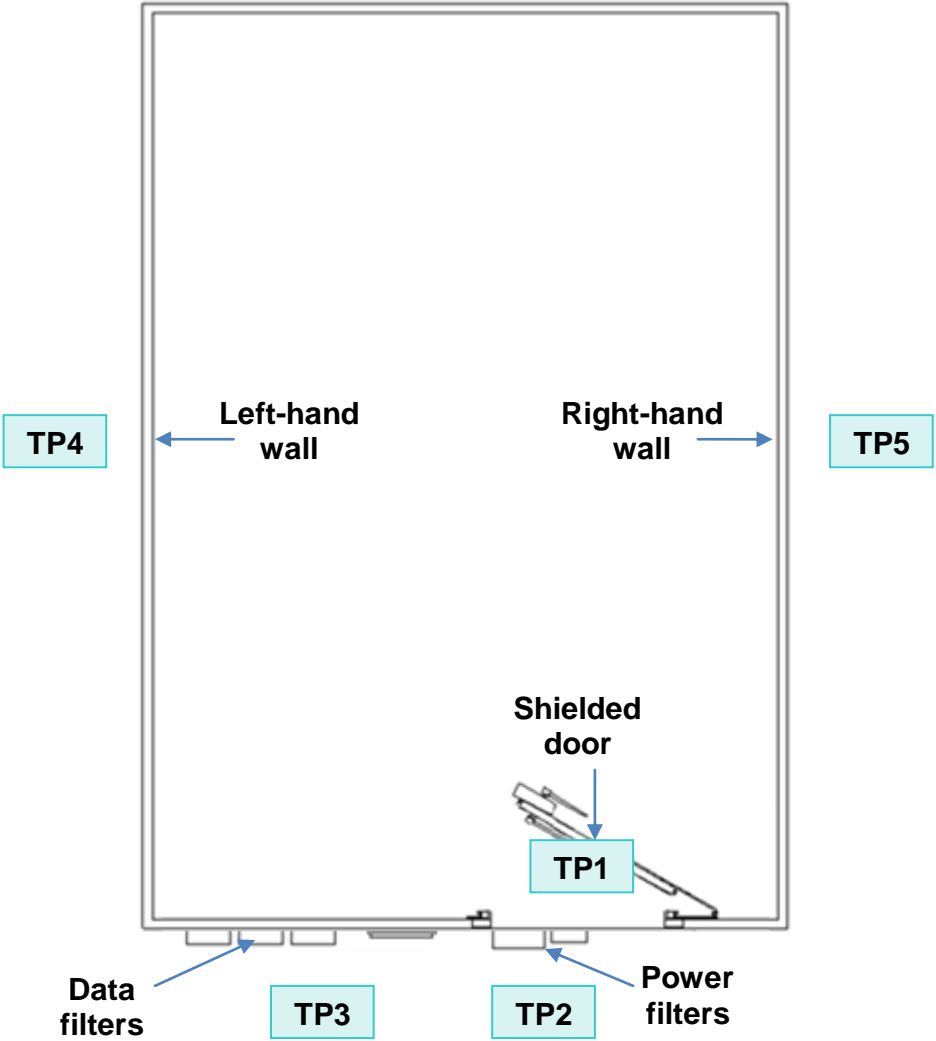

**Figure 3:** Shielded room test points

## 6. Measurement results

Measurements were performed at test points detailed in Figure 3 and the results of the shielding attenuation measurements are recorded in this section.

Where the shielding attenuation was so high that no received signal could be measured, then the dynamic range of the measurement system is given to indicate a minimum level of shielding effectiveness. This is indicated in the results as "> X dB", where X is the dynamic range of the measurement system.

The shielded effectiveness results are given in the tables below. The recorded values show the minimum level that was measured at each position.

All measurement results are expressed in dB.

| Frequency | TP1 Door | TP2 Power filters | TP3 Data filters | TP4 Left hand wall | TP5 Right hand wall |
|-----------|----------|-------------------|------------------|---------------------|----------------------|
| 1 GHz | 119 | 111 | 105 | 122 | 125 |
| 5 GHz | 111 | 102 | 95 | >115 | >115 |
| 10 GHz | 109 | 104 | 92 | 110 | >115 |

# 7. Photographs of the measurements

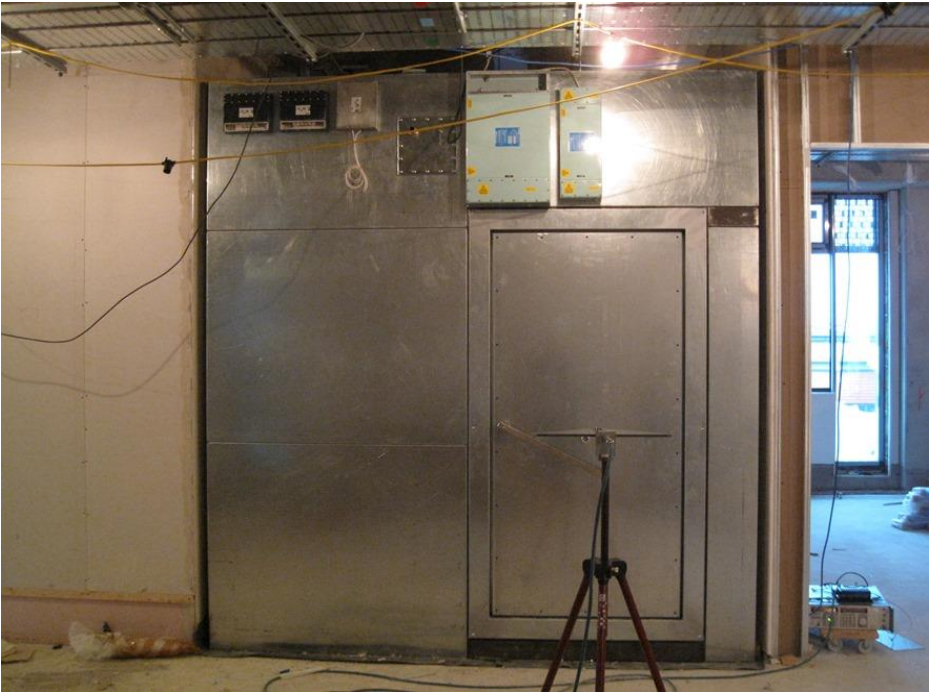

**Photograph 1:** Shielding measurement at 1 GHz (Test point 1)

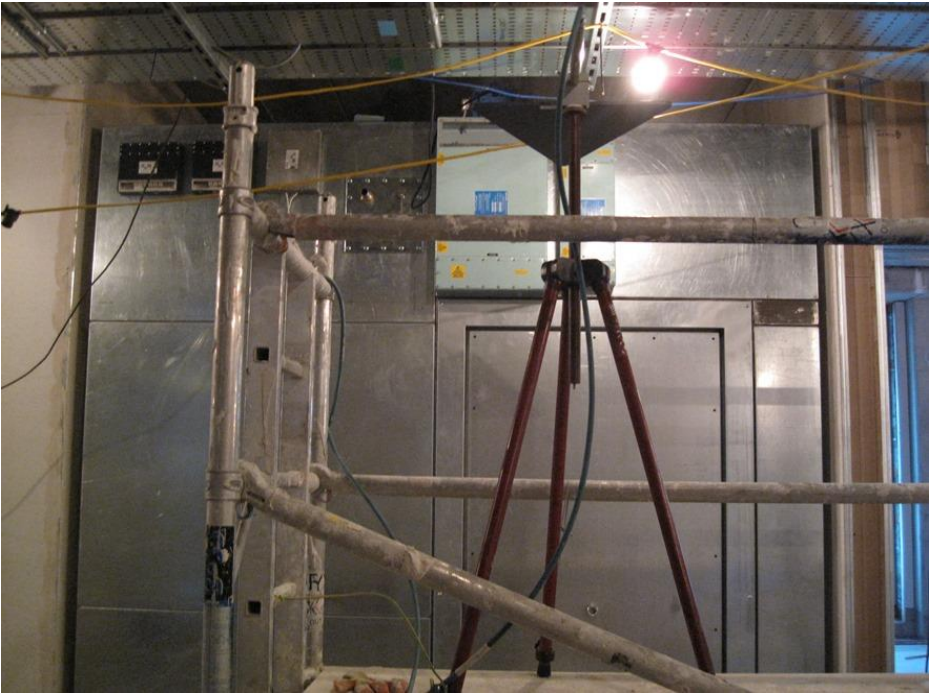

**Photograph 2:** Shielding measurement at 1 GHz (Test point 2)

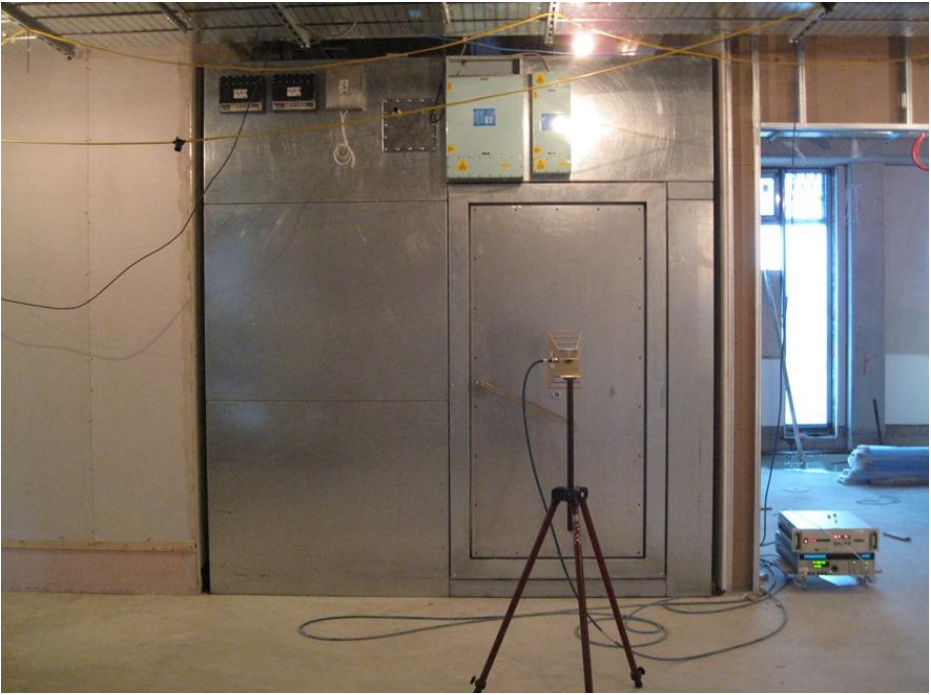

**Photograph 3:** Shielding measurement at 5 & 10 GHz (Test point 1)

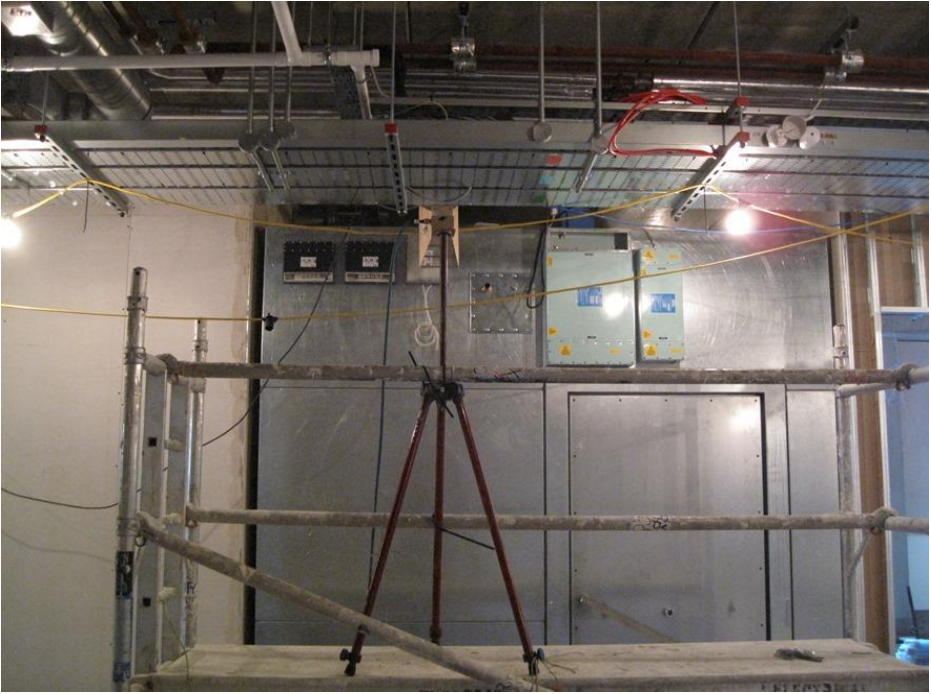

**Photograph 4:** Shielding measurement at 5 & 10 GHz (Test point 3)

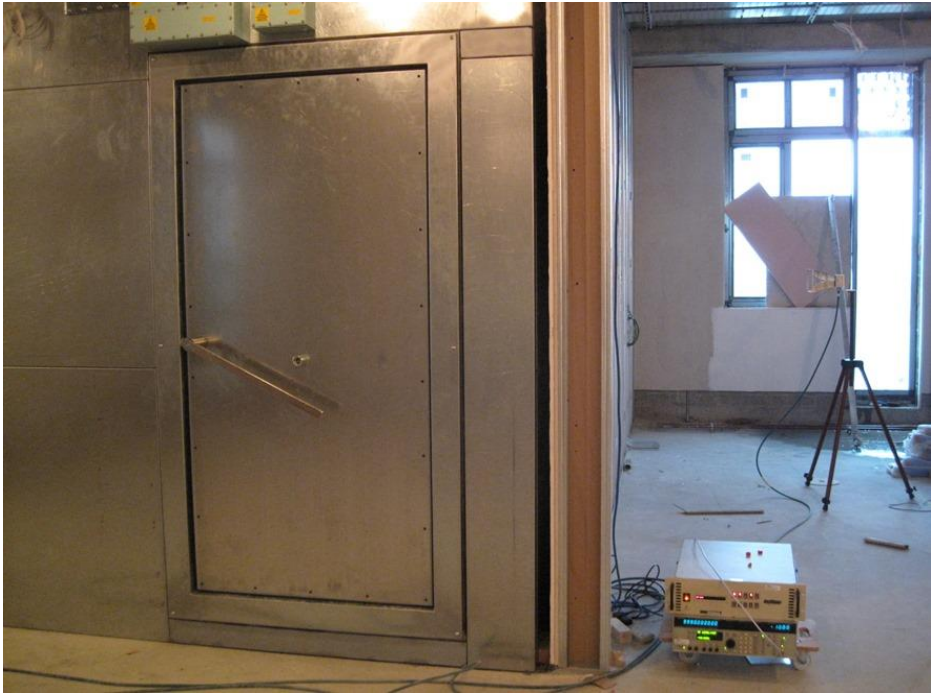

**Photograph 5:** Shielding measurement at 5 & 10 GHz (Test point 5)

# 8. References

[1]     EN 50147-1:1996,   Anechoic chambers. Part 1: Shield attenuation measurement

# Engagement Experiment Form

Please answer all questions required. Hit submit when completed.

* Required

1. **Participant Code** *

   _____

2. **Gender** *

   *Check all that apply.*

   ☐ Male

   ☐ Female

3. **Age** *

   *Mark only one oval.*

   ◯ 18-20

   ◯ 21-23

   ◯ 24 and over

4. **What hand do you use for writing?** *

   *Mark only one oval.*

   ◯ Left

   ◯ Right

5. **What hand do you hold your knife in when you are eating?** *

   *Mark only one oval.*

   ◯ Left

   ◯ Right

6. **Are you a native English speaker?** *

   *Mark only one oval.*

   ◯ Yes

   ◯ No

7. **If no, do you sufficient mastery to study in Ireland? (as measured by a recognised English examination eg. score of 6.0 UG or 6.5 PG IELTS examination)**

   *Mark only one oval.*

   ◯ Yes

   ◯ No

8. **Would you say you are a morning or night person?** *
   *Mark only one oval.*

   ◯ Morning

   ◯ Night

   ◯ No preference

9. **Do you wear prescription glasses/contacts?** *
   *Mark only one oval.*

   ◯ Yes

   ◯ No

10. **Have you had previous programming experience?** *
    *Mark only one oval.*

    ◯ Yes

    ◯ No

11. **If yes, how would you rate your level in Java** *
    *Mark only one oval.*

    |           | 1 | 2 | 3 | 4 | 5 |           |
    |-----------|---|---|---|---|---|-----------|
    | Very poor | ◯ | ◯ | ◯ | ◯ | ◯ | Very good |

12. **How many years programming experience do you have?** *
    *Mark only one oval.*

    ◯ 0-4 years

    ◯ 4-8 years

    ◯ 8+ years

13. **How often do you program?** *
    *Mark only one oval.*

    |            | 1 | 2 | 3 | 4 | 5 |       |
    |------------|---|---|---|---|---|-------|
    | Not at all | ◯ | ◯ | ◯ | ◯ | ◯ | Daily |

14. **What is your highest qualification?** *
    *Mark only one oval.*

    ◯ Leaving Certificate

    ◯ Bachelors Degree

    ◯ Masters

    ◯ PhD.

Powered by
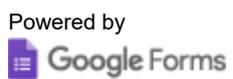


\end{document}